\documentclass{siamltex}
\usepackage[numbers]{natbib}
\usepackage{amsmath}
\usepackage{amscd}
\usepackage{amssymb}
\usepackage{graphicx}
\usepackage{subfig}
\usepackage{url}
\usepackage{color}
\usepackage{listings}
\usepackage{booktabs}
\usepackage{caption}
\usepackage{textcomp}
\definecolor{dark-gray}{gray}{0.3}
\definecolor{mauve}{rgb}{0.58,0,0.82}

\newcommand{\dx}{\,\mathrm{d}x}
\newcommand{\dt}{\,\mathrm{d}t}

\title{Automated derivation of the adjoint of high-level transient finite element programs}

\author{
  P. E. Farrell\thanks{Department of Earth Science and
    Engineering, Imperial College London, London, UK.
    Center for Biomedical Computing, Simula Research Laboratory, Oslo, Norway
    (\texttt{patrick.farrell@imperial.ac.uk}).}
  \and
  D. A. Ham\thanks{Department of Computing,
    Imperial College London, London, UK. Grantham Institute for
    Climate Change, Imperial College London, London, UK
    (\texttt{david.ham@imperial.ac.uk}).}
  \and
  S. W. Funke\thanks{Department of Earth Science and Engineering,
    Imperial College London, London, UK. Grantham Institute for
    Climate Change, Imperial College London, London, UK
    (\texttt{s.funke09@imperial.ac.uk}).}
  \and
  M. E. Rognes\thanks{Center for Biomedical Computing, Simula Research
    Laboratory, Oslo, Norway (\texttt{meg@simula.no}).
    This research is funded by NERC grant NE/I001360/1, EPSRC grant
    EP/I00405X/1, the Grantham Institute for Climate Change, a Fujitsu
    CASE studentship, a Center of Excellence grant from the Research
    Council of Norway to the Center for Biomedical Computing at Simula
    Research Laboratory, and an Outstanding Young Investigator grant
    from the Research Council of Norway, NFR 180450.}  }

\begin{document}

\maketitle

\begin{abstract}
In this paper we demonstrate a new technique for deriving discrete
adjoint and tangent linear models of finite element models. The
technique is significantly more efficient and automatic than standard
algorithmic differentiation techniques. The approach relies on a
high-level symbolic representation of the forward problem.  In contrast
to developing a model directly in Fortran or C++, high-level systems
allow the developer to express the variational problems to be solved in
near-mathematical notation. As such, these systems have a key advantage:
since the mathematical structure of the problem is preserved, they are
more amenable to automated analysis and manipulation.  The framework
introduced here is implemented in a freely available software package
named dolfin-adjoint, based on the FEniCS Project.  Our approach to
automated adjoint derivation relies on run-time annotation of the
temporal structure of the model, and employs the FEniCS finite element
form compiler to automatically generate the low-level code for the
derived models.  The approach requires only trivial changes to a large
class of forward models, including complicated time-dependent nonlinear
models. The adjoint model automatically employs optimal checkpointing
schemes to mitigate storage requirements for nonlinear models, without
any user management or intervention. Furthermore, both the tangent
linear and adjoint models naturally work in parallel, without any need
to differentiate through calls to MPI or to parse OpenMP directives. The
generality, applicability and efficiency of the approach are
demonstrated with examples from a wide range of scientific applications.
\end{abstract}



\section{Introduction}

Adjoint models are key ingredients in many algorithms of computational
science, such as in parameter identification \cite{navon1998},
sensitivity analysis~\cite{cacuci1981}, data
assimilation~\cite{ledimet1986}, optimal control~\cite{lions1971},
adaptive observations~\cite{palmer1998}, predictability
analysis~\cite{lorenz1965}, and error
estimation~\cite{becker2001}. While deriving the adjoint model
associated with a linear stationary forward model is straightforward,
the development and implementation of adjoint models for nonlinear or
time-dependent forward models is notoriously difficult, for several
reasons. First, each nonlinear operator of the forward model must be
differentiated, which can be difficult for complex models. Second, the
control flow of the adjoint model runs backwards, from the final time
to the initial time, and requires access to the solution variables
computed during the forward run if the forward problem is
nonlinear. Since it is generally impractical for physically relevant
simulations to store all variables during the forward run, the adjoint
model developer must implement some checkpointing scheme that balances
recomputation and storage~\cite{griewank1992}. The control flow of
such a checkpointing scheme must alternate between the solution of
forward variables and adjoint variables, and is thus highly nontrivial
to implement by hand on a large and complex code. For parallel
computations, these difficulties are magnified by the fact that the
control flow of parallel communications reverses in the adjoint solve:
forward sends become adjoint receives, and forward receives become
adjoint sends~\cite{schanen2010}.

\begin{figure}
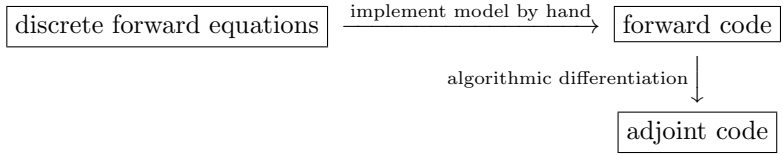

\begin{equation*}
  \begin{CD}
    \fbox{\textrm{discrete forward equations}}
    @>\textrm{implement model by hand}>>
    \fbox{\textrm{forward code}}\\
    @.
    @V\textrm{algorithmic differentiation}VV \\
    @.
    \fbox{\textrm{adjoint code}}\\
  \end{CD}
\end{equation*}
  \caption{The traditional approach to developing adjoint models. The forward model
  is implemented by hand, and its adjoint derived either by hand or with the assistance
  of an algorithmic differentiation tool.}
  \label{fig:traditional_approach}
\end{figure}

The traditional approach to model development is to implement the forward code
by hand in a low-level language (typically Fortran or C++). While this allows
the programmer a high degree of control over each memory access and floating
point operation, implementing these codes usually takes a large amount of time,
and the mathematical structure of the problem to be solved is irretrievably
interwoven with implementation details of how the solution is to be achieved.
Then the adjoint code is produced, either by hand or with the assistance of an
algorithmic differentiation (AD) tool (figure \ref{fig:traditional_approach}).
Such AD tools take as input a forward model written in a low-level language, and
derive the associated discrete adjoint model, through some combination of
source-to-source transformations and operator overloading.  However, this
process requires expert knowledge of both the tool and the model to be
differentiated \cite[pg. \emph{xii}]{naumann2011}.  The root cause of the
difficulty which AD tools have is that they operate on low-level code in which
implementation details and mathematics are inseparable and therefore must both
be differentiated: AD tools must concern themselves with matters such as memory
allocations, pointer analyses, I/O, and parallel communications (e.g. MPI or
OpenMP).

A variant of this approach is to selectively apply AD to small
sections of the model, and then to connect and arrange these
differentiated routines by hand to assemble the discrete adjoint
equations~\cite{giles2005, coleman1998, marta2007}.  This approach
attempts to re-introduce as much as possible of the distinction
between mathematics and implementation; however, it requires even more
expertise than a na\"ive black-box application of AD.

Algorithmic differentiation treats models as a sequence of elementary
instructions, where an instruction is typically a native operation of
the programming language such as addition, multiplication or
exponentiation.  Instead, we consider a new, higher-level abstraction
for developing discrete adjoint models: to treat the model as a
sequence of equation solves. This offers an alternative approach to
the development of discrete adjoint models, and is implemented in an
open-source software library called libadjoint.

When libadjoint is applied to a low-level forward code, the developer must
\emph{annotate} the forward model. This involves embedding calls to the
libadjoint library that record the temporal structure of the equations as they
are solved. The recorded information is analogous to a tape in AD, but at a
higher level of abstraction. Using this information, libadjoint can symbolically
manipulate the annotated system to derive the structure of the adjoint equations
or the tangent linear equations. If the adjoint developer further supplies
\emph{callback functions} for each operator that features in the annotation and
any necessary derivatives, libadjoint can assemble each adjoint or tangent
linear equation as required. The library therefore relieves the developer of
deriving the adjoint equations, managing the complex life cycles of forward and
adjoint variables, and implementing a checkpointing scheme. With this strategy,
the task of developing the adjoint model (which requires significant expertise)
is replaced with the tasks of describing and modularising it (which are usually
much more straightforward).

The aim of this work is to apply libadjoint to automatically derive
the adjoint of models written in a high-level finite element system.
In such systems, the discrete variational formulation is expressed in
code which closely mimics mathematical notation.  The low-level
details of finite element assembly and numerical linear algebra are
delegated to the system itself. For example, in the Sundance C++
library, the assembly is achieved by runtime Fr\'echet differentiation
of the specified variational form~\cite{long2010}. In the FEniCS
environment, the variational form is passed to a dedicated finite
element form compiler, which generates low-level code for its
assembly~\cite{kirby2006,logg2010a,logg2011a}. By exploiting
optimisations that are impractical to perform by hand, such systems
can generate very efficient implementations~\cite{kirby2007b,
  olgaard2010, markall2012}. Moreover, a major advantage of this clean
separation between mathematical intention and computer implementation
is that it enables the automatic mathematical analysis and
manipulation of the variational form. The absence of this separation
in low-level models inhibits automated analysis at the level of
variational forms. In addition, high-level systems can often provide
automated variational form differentiation/linearisation
capabilities~\cite{alnaes2012,alnaes2011,long2010,prudhomme2006}, which can be of
particular interest for adjoint models.

\begin{figure}
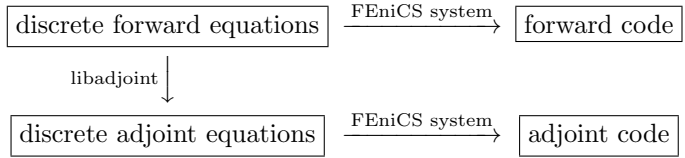

\begin{equation*}
  \begin{CD}
    \fbox{\textrm{discrete forward equations}}
    @>\textrm{FEniCS system}>>
    \fbox{\textrm{forward code}}\\
    @V\textrm{libadjoint}VV
    \\
    \fbox{\textrm{discrete adjoint equations}}
    @>\textrm{FEniCS system}>>
    \fbox{\textrm{adjoint code}}\\
  \end{CD}
\end{equation*}
  \caption{The approach to adjoint model development advocated in this
    paper. The user specifies the discrete forward equations in a
    high-level language similar to mathematical notation; the discrete
    forward equations are explicitly represented in memory in the UFL
    format~\cite{alnaes2012,alnaes2011}.  libadjoint automatically derives the
    corresponding in-memory representation of the discrete adjoint
    equations, from the in-memory representation of the forward
    problem.  Both the forward and adjoint equations are then passed
    to the FEniCS system, which automatically generates and executes
    the code necessary to compute the forward and adjoint solutions.}
  \label{fig:our_approach}
\end{figure}

The main contribution of this paper is a new framework for the
automated derivation of the discrete adjoint and tangent linear models
of forward models implemented in the FEniCS software environment. The
strategy is illustrated in figure~\ref{fig:our_approach}. In FEniCS
model code, each equation solve may be naturally expressed as a single
function call. This matches the basic abstraction of libadjoint
exactly, and therefore makes the integration of FEniCS and libadjoint
particularly straightforward.  Consequently, for a large class of
forward models implemented in FEniCS, libadjoint can automatically
derive the discrete adjoint model with only minor additions to the
code. The necessary derivative terms are automatically computed using
the form linearisation capabilities of UFL~\cite{alnaes2012,alnaes2011}.  The
adjoint equations derived by libadjoint are themselves valid FEniCS
input, and the low-level adjoint code is generated using the same
finite element form compiler as the forward model. As a result, the
derived adjoint models approach optimal efficiency, automatically
employ optimal checkpointing schemes, and inherit parallel support
from the forward models.

This paper is organised as follows. In section~\ref{sec:libadjoint},
we describe the fundamental abstraction underlying libadjoint and give
a brief review of libadjoint. In section~\ref{sec:fenics}, we outline
the implementation of transient finite element forward models in the
FEniCS framework.  The integration of FEniCS and libadjoint is
implemented in a new software framework called dolfin-adjoint,
which is described in section~\ref{sec:dolfin_adjoint}. The advantages
and limitations of the approach are discussed in
section~\ref{sec:discussion}.  Numerical examples drawn from a wide
range of scientific applications are presented in
section~\ref{sec:examples}, before we make some concluding remarks in
section~\ref{sec:conclusion}.

\section{The fundamental abstraction of libadjoint} \label{sec:libadjoint}
In this section, we detail the basic abstraction upon which libadjoint
is based. This abstraction is to treat the model as a sequence of
equation solves. This abstraction
applies to both stationary and time-dependent systems of partial
differential equations, and to both linear and nonlinear systems.

\subsection{Mathematical framework}

We consider systems of discretised partial differential equations
expressed in the fundamental abstract form
\begin{equation}
  \label{eqn:libadjoint_form}
  A(u) u = b(u),
\end{equation}
where $u$ is the vector of all prognostic variables, $b(u)$ is the
source term, and $A(u)$ the entire discretisation matrix. In the
time-dependent case, $u$ is a block-structured vector containing all
the values of the unknowns at all the time levels, $A$ is a matrix
with a lower-triangular block structure containing all of the
operators featuring in the forward model, and $b$ is a
block-structured vector containing all of the right-hand side terms
for all of the equations solved in the forward model. The
block-lower-triangular structure of $A$ is a consequence of the
forward propagation of information through time: later values depend
on earlier values, but not vice versa.

It is to be emphasised that writing the model in this
format~\eqref{eqn:libadjoint_form} does not imply that the whole of
$A$ is ever assembled at once, or the whole of $u$ stored in
memory. For instance, the forward solver will typically assemble one
block-row of $A$, solve it for a block-component of $u$, forget as
much as possible, and step forward in time.

Let $m$ be some parameter upon which the forward equations depend.
For example, $m$ could be a boundary condition, initial condition, or
coefficient appearing in the equations.  The tangent linear model
associated with~\eqref{eqn:libadjoint_form} is then given by
\begin{equation} \label{eqn:tangent_linear}
(A + G - R) \frac{du}{dm} = - \frac{\partial F}{\partial m},
\end{equation}
where
\begin{align}
G \equiv & \ \frac{\partial A}{\partial u} u, \label{eqn:matrix_derivative} \\
R \equiv & \ \frac{\partial b}{\partial u}, \label{eqn:R_matrix}
\end{align}
and
\begin{align}
F \equiv & \ A(u)u - b(u).
\end{align}
The unknown in~\eqref{eqn:tangent_linear} is the ${du}/{dm}$ matrix,
the Jacobian of the solution $u$ with respect to the parameters $m$.
The $G$ matrix of~\eqref{eqn:matrix_derivative} arises because of the
nonlinear dependency of the operator $A$. $A$ is a matrix (i.e. a rank-2 tensor), so
differentiating it with respect to $u$ yields a rank-3 tensor; the
following contraction with $u$ over the middle index reduces the rank
again to 2. More precisely in index notation we have,
\begin{align} \label{eqn:G_index}
  G_{ik} \equiv & \ \sum_j \frac{\partial A_{ij}}{\partial u_k} u_j.
\end{align}
The $R$ matrix of~\eqref{eqn:R_matrix} is the Jacobian of the
right-hand side $b$ with respect to the solution $u$. Written in index
notation,
\begin{align}
R_{ij} \equiv \frac{\partial b_i}{\partial u_j}.
\end{align}

Let $J$ be some functional of the solution $u$. $J$ is a function that
takes in the system state, and returns a single scalar diagnostic. For
example, in aeronautical design, $J$ may be the drag coefficient
associated with a wing; in meteorology, $J$ may be the weighted misfit
between observations of the atmosphere and model results. The adjoint
model associated with~\eqref{eqn:libadjoint_form} is given by
\begin{equation} \label{eqn:adjoint_eqn}
(A + G - R)^* z = \frac{\partial J}{\partial u},
\end{equation}
where $z$ is the adjoint solution associated with
$J$, and $^*$ denotes the conjugate transpose.

To make matters more concrete, consider the following example. Suppose
the forward model approximately solves the time-dependent viscous
Burgers' equation
\begin{equation}\label{eqn:burgerscont}
\frac{\partial u}{\partial t} + u \cdot \nabla u - \nabla^2 u = f,
\end{equation}
for the velocity $u$, subject to some suitable boundary conditions and
a supplied initial condition $u(0) = g$, with source term
$f$. For simplicity, suppose the model linearises the nonlinear advective term around the solution of
the previous timestep (other choices are also possible). Discretising with the Galerkin finite element method in space and
the forward Euler method in time yields the
timestep iteration
\begin{equation}
  \begin{split}
  \label{eqn:burgers}
u_0 \leftarrow& g \\
M u_{n+1}       \leftarrow& (M - \Delta t V\left(u_n\right) - \Delta t D)u_n + \Delta t f_n,
  \end{split}
\end{equation}
where $n$ is the timelevel, $\Delta t$ is the timestep, $M$ is the
mass matrix, $D$ is the diffusion matrix, $V(u)$ is the advection
matrix assembled at a given velocity $u$. For brevity, define
\begin{equation}
T(\cdot) \equiv \Delta t V(\cdot) + \Delta t D - M.
\end{equation}
The system \eqref{eqn:burgers} can be cast
into the form of~\eqref{eqn:libadjoint_form} by writing $N$ timestep
iterations as
\begin{equation} \label{eqn:burgers_A}
\begin{pmatrix}
I \\
T(u_0) & M \\
& T\left(u_1\right) & M \\
  & & \ddots & \ddots \\
  & & & T\left(u_{N-1}\right) & M
\end{pmatrix}
\begin{pmatrix}
u_0 \\
u_1 \\
u_2 \\
\vdots \\
u_N
\end{pmatrix} =
\begin{pmatrix}
g \\
\Delta t f_0 \\
\Delta t f_1 \\
\vdots \\
\Delta t f_{N-1}
\end{pmatrix}.
\end{equation}
Using~\eqref{eqn:adjoint_eqn}, the adjoint system is given by
\begin{equation} \label{eqn:burgers_A_adjoint}
\begin{pmatrix}
I^* & \left(T(u_0) + \frac{\partial T(u_0)}{\partial u_0} u_0\right)^* &   &  & \\
    & M^*  &  \left( T(u_1) + \frac{\partial T(u_1)}{\partial u_1} u_1\right)^*  & & \\
    &      &  \ddots & \ddots & \\
    &      & & \ddots & \ddots \\
    &      & & & M^*
\end{pmatrix}
\begin{pmatrix}
z_0 \vphantom{\left(T(u_0) + \frac{\partial T(u_0)}{\partial u_0} u_0\right)^* } \\
z_1 \vphantom{\left( T(u_1) + \frac{\partial T(u_1)}{\partial u_1} u_1\right)^*} \\
\vdots \\
\vdots \\
z_N
\end{pmatrix}
=
\frac{\partial J}{\partial u},
\end{equation}
where the contraction of the derivative of a matrix with a vector is defined in
\eqref{eqn:G_index}.

The adjoint system reverses the temporal flow of information: where
the forward and tangent linear models are block-lower-triangular, the
adjoint model is block-upper-triangular, as visible
in~\eqref{eqn:burgers_A_adjoint}. The adjoint system is therefore
typically solved by backward substitution: the adjoint variable
associated with the end of time is solved for first, and then the
solution proceeds backwards in time.

At its heart, libadjoint takes models cast in the form of
\eqref{eqn:libadjoint_form}, and derives, assembles and solves the
tangent linear \eqref{eqn:tangent_linear} and adjoint
\eqref{eqn:adjoint_eqn} systems block-row by block-row.

\subsection{Libadjoint}

Applying libadjoint to a model breaks down into two main tasks.
The automation of these two tasks by dolfin-adjoint, the software
package presented in this paper,
is described in section~\ref{sec:dolfin_adjoint}.

The first task, referred to as \emph{annotation}, is to describe the
forward code in the form of the fundamental
abstraction~\eqref{eqn:libadjoint_form}.  By describing the forward
code in this form, libadjoint automates the reasoning necessary to
derive the adjoint~\eqref{eqn:adjoint_eqn} and tangent
linear~\eqref{eqn:tangent_linear} systems.  As each equation is
solved, the necessary semantic information about that equation is
recorded. In particular, each equation records the variable solved
for, the operators (matrices) that feature in the equation, and any
dependencies these operators have on previous variables.  This
annotation is effected by making calls to library functions offered by
the libadjoint API. In a low-level code, these calls must be inserted
manually by the model developer, as the high-level semantic structure
will have been obscured in the process of implementing the model.  The
annotation enables libadjoint to symbolically derive the structure of
the discrete adjoint and tangent linear systems; however, without
further information it cannot assemble the actual adjoint or tangent
linear equations.  The operators in the annotation are mere abstract
handles.

The second task is to supply libadjoint with function callbacks for
the operators that feature in the annotation. If these operators
depend on previously computed variables, their derivatives must also
be supplied; the code for these derivative callbacks may be written by
hand, or may be generated with an AD tool. With these callbacks, the
derived equations may be automatically assembled. By modularising the
forward model in this manner, libadjoint can drive the assembly of the
forward, tangent linear and adjoint models, by calling the appropriate
callbacks in the correct sequence. In a low-level code, these
callbacks must be written manually. This can be a significant burden
if the original forward code is poorly modularised.

The use of libadjoint offers several advantages. It can be applied to
models for which black-box AD is intractable, and
gains the speed and efficiency benefits of applying AD
judiciously~\cite{giles2005, coleman1998, marta2007}. Using libadjoint
makes development systematic: each incremental step in its application
may be rigorously verified. As libadjoint internally derives a
symbolic representation of the discrete adjoint equations, it can
compute when a forward or adjoint variable is no longer necessary, and
thus the model developer is relieved of the management of variable
deallocation.  Furthermore, libadjoint can automatically check the
consistency of the adjoint computed with the original forward model;
this check greatly improves the maintainability of adjoint codes, as
developers can be immediately notified when a change to the forward
model is not mirrored in the adjoint model.  Finally, as libadjoint
has sufficient information to re-assemble the forward equations, it is
possible to implement checkpointing schemes entirely within the
library itself.  Checkpointing is a crucial feature for the efficient
implementation of the adjoint of a time-dependent nonlinear model, but
its implementation can be prohibitively difficult. The availability of
optimal checkpointing schemes within libadjoint is a significant
advantage.

\section{The FEniCS system} \label{sec:fenics}

The FEniCS Project is a collection of software
components for automating the solution of differential
equations~\cite{logg2011, logg2007a}. These components include the
Unified Form Language (UFL)~\cite{alnaes2012,alnaes2011}, the FEniCS Form
Compiler (FFC)~\cite{kirby2006}, and DOLFIN~\cite{logg2010a,
  logg2011a}. In the following, we only briefly outline the FEniCS
pipeline, and refer the reader to the aforementioned references for
more information.

One of the key features of the FEniCS components is the use of code
generation, and in particular domain-specific code generation for
finite element variational formulations: the user specifies the
discrete variational problem to be solved in the domain-specific
language UFL, the syntax of which mimics and encodes the mathematical
formulation of the problem. Based on this high-level formulation, a
special-purpose finite element form compiler generates optimised
low-level C++ code for the evaluation of local element tensors.  The
generated code is then used by DOLFIN to perform the global assembly
and numerical solution. DOLFIN also provides the underlying data
structures such as meshes, function spaces, boundary conditions and
function values.  DOLFIN provides both a C++ and a Python
interface. For the C++ interface, the UFL specification and the form
compilation must take place off-line, and the generated code is
explicitly included by the user. In the Python interface, the
functionality is seamlessly integrated by way of runtime just-in-time
compilation. The approach taken by dolfin-adjoint is only applicable to models
written using the Python interface, as only the Python interface has
runtime access to the symbolic description of the forward model in
UFL format.

DOLFIN abstracts the spatial discretisation problem, but not the
temporal discretisation problem.  Transient DOLFIN-based solvers
typically consist of a hand-written temporal loop in which one or more
discrete variational (finite element) problems are solved with DOLFIN
in each iteration. An overview of common operations relevant to
dolfin-adjoint is given in table~\ref{tab:dolfin:func}.
\begin{table}
  \begin{center}
    \begin{tabular}[ht]{ll}
      \toprule
      DOLFIN function signature & Short description \\
      \midrule
      \texttt{solve(lhs == rhs, u, bcs)}
      & Solve variational problem \\
      \texttt{u.assign(u\_)}
      & Copy function u\_ to u \\
      \texttt{assemble(a)}
      & Assemble variational form \\
      \texttt{bc.apply(A)}
      & Apply a boundary condition to a matrix\\
      \texttt{solve(A, x, b)}
      & Solve linear system \\
      \texttt{project(u, V)}
      & Project $u$ onto the function space $V$ \\
      \bottomrule
    \end{tabular}
    \caption{Important DOLFIN statements relevant to dolfin-adjoint.}
    \label{tab:dolfin:func}
  \end{center}
\end{table}

At the highest level of abstraction, a DOLFIN model developer can
define the variational problem for each timestep in terms of the
unknown function, the left and right hand side forms and boundary
conditions, and then call the DOLFIN \texttt{solve} function.  If the
variational problem is linear (represented by a left-hand side
bilinear form, and a right-hand side linear form, \texttt{a == L}),
the linear system of equations is assembled and solved under the
hood. If the variational problem is nonlinear (represented by a
left-hand side rank-1 form and zero right-hand side, \texttt{F
  == 0}), a Newton iteration is invoked in which the Jacobian of the
variational form \texttt{F} is derived automatically, and the linear
system in each Newton iteration assembled and solved until the
iteration has converged. The derivation of the Jacobian employs the
algorithmic differentiation algorithms of UFL: because the form is
represented in a high-level abstraction, the symbolic differentiation
of the form is straightforward~\cite[\S17.5.2, \S17.7]{alnaes2011}.
When moving from one timestep to the next, the \texttt{assign}
function is typically used to update the previous function with the
new value. A sample solver demonstrating this type of usage, solving
the nonlinear Burgers' equation, is listed in
figure~\ref{fig:dolfin_burgers}.

However, DOLFIN also supports more prescriptive programming models, in
which explicit calls are employed to assemble matrices and solve the
resulting linear systems. If the variational problem is linear and the
left-hand side is constant, the assembly of the stiffness matrix may
occur outside the temporal loop, and the matrix reused. The solution
of the linear systems may also be further controlled by specifying
direct LU or Krylov solvers, or even matrix-free solvers.

DOLFIN's abstraction of a transient problem as an explicit sequence of
variational problems exactly matches the fundamental abstraction of
libadjoint; this matching of abstractions is the basis of the work
presented here.

\section{Applying libadjoint to DOLFIN} \label{sec:dolfin_adjoint}
\begin{figure}[ht]
  \centering
  \input{burgers}
  \caption{DOLFIN code for a simple discretisation of the Burgers
    equation~\eqref{eqn:burgerscont} with
    dolfin-adjoint\ annotations. The
    dolfin-adjoint\ module overloads the existing
    \texttt{solve}\ and \texttt{assign} functions (indicated in red), and allows the user
    to specify names of \texttt{Function}s for convenience. The only
    change to the code body is the introduction of a call to
    \texttt{adj\_inc\_timestep} to indicate to libadjoint that a new
    timestep is commencing (indicated in blue).}
  \label{fig:dolfin_burgers}
\end{figure}

In this section, we present the internal details of how
dolfin-adjoint integrates DOLFIN with libadjoint. This
includes the annotation of the forward model execution, the recording
of any necessary values, and the generation and registration of
callback functions. All of these processes happen automatically,
without any intervention by the model developer.

\subsection{Annotations} \label{sec:annotations}
The basic mechanism of automatically annotating DOLFIN models employed
here is to overload the DOLFIN functions that change the values of
variables. The overloaded versions annotate the event and then pass
control to the original DOLFIN functions. All of the functions listed
in table~\ref{tab:dolfin:func} are annotated.
Figure~\ref{fig:dolfin_burgers}\ presents a Burgers' equation model
modified for use with dolfin-adjoint. The modifications and
overloaded functions are highlighted. This demonstrates that only
minimal source changes are required.

In a low-level model, the information that libadjoint needs to record
is not explicitly represented as data in the code; therefore,
annotation has to be at least partially done by hand, as the
programmer must supply the information instead. By contrast, in the
Python interface to DOLFIN, the equation to be assembled is explicitly
represented as data at runtime; DOLFIN's \texttt{solve} function takes
as input the variational form of the equation to be solved, and so all
information necessary for the annotation is available during the
\texttt{solve} function. This makes the automatic annotation
possible. By contrast, when using the C++ interface to DOLFIN, the
code generation happens offline, and the equations are not explicitly
represented as data at runtime; it is for this reason that the
automated runtime derivation of the adjoint model in the manner
described here is not possible.  Therefore, dolfin-adjoint
only supports the use of the Python interface to DOLFIN.

When called, the adjoint-aware \texttt{solve} function inspects the
left- and right-hand sides of the provided equation to determine the
information needed by libadjoint. This includes the variable being
solved for, the operators which feature in the equation, and their
dependencies on values that were previously computed. UFL supports the
interrogation of forms to automatically extract the necessary
information.  This information is then registered with libadjoint
using the relevant API calls.  When an overloaded call completes,
dolfin-adjoint will record the resulting value if this is
required for later use in the adjoint computation.

For nonlinear solves expressed in the form $F(u) = 0$, where $F$ is a
rank-1 (vector) form, the adjoint code annotates the equivalent
equation
\begin{equation} \label{eqn:nonlinear_annotation}
Iu = Iu - F(u),
\end{equation}
where $I$ is the identity matrix associated with the function space of
$u$, so that it can be cast into the form of a row
of~\eqref{eqn:libadjoint_form}.  While it would be possible to
annotate each linear solve in the Newton iteration, this would be
inefficient: with this method, the adjoint run would have to rewind
through each iteration of the nonlinear solver, whereas by annotating
in the manner of~\eqref{eqn:nonlinear_annotation}, only one linearised
solve is necessary. This is akin to the method suggested
in~\cite{gilbert1992}, which computes the derivative of a Newton
iteration in one step by linearising about the computed forward
solution.

In the cases where a developer of DOLFIN models pre-assembles forms
into tensors using the \texttt{assemble} function, only low-level
matrices and vectors are given as inputs to the \texttt{solve} call;
the semantic information about the forms is no longer available.  This
problem is resolved by supplying an overloaded \texttt{assemble}
function which associates the form to be assembled with the assembled
tensor. When the matrix-vector version of \texttt{solve} is called, it
uses this association to recover the forms involved in the equation,
and annotates as described above.

The correctness of the adjoint relies on the correctness of the
annotation. If the annotation does not exactly record the structure of
the forward model, the gradient computed using the adjoint will be
inconsistent. For example, the annotation would become inconsistent if
the model accesses the underlying \texttt{.vector()} of a
\texttt{Function} and changes the values by hand; this would not be
recorded by libadjoint. This restriction is mitigated somewhat by the
fact that the replay feature of libadjoint can be used to check the
correctness of the annotation. As will be discussed in
section~\ref{sec:checkpointing}, libadjoint has sufficient information
to replay the forward equations. This can be used to automatically
check the correctness of the annotation, by replaying the annotation
and comparing each value to that computed during the original model
run.

\subsection{Callbacks}
In the dolfin-adjoint code, the registration of callbacks
occurs at the same time as the annotation, in the overloaded function
calls. For each operator in the equations registered, Python functions
are generated by dolfin-adjoint at runtime, and these are
associated with the operator through the relevant libadjoint API
calls.  Depending on the precise details of the operator, libadjoint's
requirements vary: if the operator is on the main diagonal of $A$
in~\eqref{eqn:libadjoint_form}, the function returns the form itself
(or its transpose, in the adjoint case), while if the operator is not
on the main diagonal, the function returns the action of the form on a
given input vector (or its transpose action). In either case, if the
form depends on previously computed solutions, the derivative of the
form with respect to these variables must be generated, along with the
associated transposes. For the computation of these transposes and
derivatives, we rely on the relevant features of UFL in which the form
is expressed, in particular on its powerful algorithmic differentiation
(AD) capabilities~\cite[\S17.5.2, \S17.7]{alnaes2011}.

The function callbacks also take references to information other than
the form, so that the exact conditions of the forward solve can be
recovered by dolfin-adjoint as necessary. In particular,
time-dependent boundary conditions and forcing terms are implemented
in DOLFIN via \texttt{Expression} classes which take in the current
time as a parameter. Before a function is defined, the current value
of every parameter is recorded, and the function then includes this
record as part of its lexical closure. When the function is called, it
restores each parameter value to the value it had when it was
created. Similarly, functions take references to any Dirichlet
boundary conditions that are applied to the equation as part of their
lexical closure, so that the associated homogeneous Dirichlet boundary
conditions may be applied to the corresponding adjoint or tangent
linear equations.

Again, because of the fact that the equation to be solved is
represented as data at runtime, the definition of the callbacks can
happen entirely automatically.  Following the registration of these
callbacks, libadjoint can compose the appropriate terms at will to
assemble the adjoint or tangent linear system corresponding
to~\eqref{eqn:libadjoint_form}.

\subsection{The dolfin-adjoint user interface}
The highest-level interface is the function
\texttt{compute\_gradient}: given a \texttt{Functional} $J$ and a
\texttt{Parameter} $m$, it computes the gradient
$\mathrm{d}J/\mathrm{d}m$ using the adjoint solution. Example usage is
given in figure~\ref{fig:adjoint_burgers}.
\begin{figure}
  \centering
  \input{adjoint_burgers.tex}
  \captionsetup{singlelinecheck=off}
  \caption[.]{Sample dolfin-adjoint user code complementing
    the Burgers model presented in figure~\ref{fig:dolfin_burgers}:
    the adjoint is generated and used to compute the gradient of the
    functional $J = \frac{1}{2} \int_{\Omega} u(T) \cdot u(T) \dx$
    with respect to the initial condition $u(0)$.}
  \label{fig:adjoint_burgers}
\end{figure}

If the user wishes to directly access the adjoint or tangent linear
solutions, the functions \texttt{compute\_adjoint} and
\texttt{compute\_tlm} are available. These functions iterate over the
adjoint and tangent linear solutions, with the tangent linear
solutions advancing in time and the adjoint solutions going backwards.
A list of important dolfin-adjoint statements is given in
table~\ref{tab:dolfin_adjoint:func}.
\begin{table}
  \begin{center}
    \begin{tabular}[ht]{ll}
      \toprule
      dolfin-adjoint statement & Short description \\
      \midrule
      \texttt{J = Functional(0.5*inner(u, u)*dx*dt[FINISH\_TIME])}
      & Functional $J = \frac{1}{2} \int_{\Omega} u(T) \cdot u(T) \dx$\\
      \texttt{J = Functional(0.5*inner(u, u)*dx*dt)}
      & Functional $J = \frac{1}{2} \int_t \int_{\Omega} u(t) \cdot u(t) \dx \dt$\\
      \texttt{m = InitialConditionParameter("Velocity")}
      & Control variable for the initial velocity \\
      \texttt{compute\_gradient(J, m)}
      & Computes the gradient $\mathrm{d}J/\mathrm{d} u_0$ \\
      \texttt{compute\_adjoint(J)}
      & Generator for the adjoint solutions \\
      \texttt{compute\_tlm(m)}
      & Generator for the tangent linear solutions \\
      \bottomrule
    \end{tabular}
    \caption{Important dolfin-adjoint statements.}
    \label{tab:dolfin_adjoint:func}
  \end{center}
\end{table}

\section{Discussion} \label{sec:discussion}
\subsection{Checkpointing} \label{sec:checkpointing}
The adjoint and tangent linear models are linearisations of the
forward model.  If the forward model is nonlinear, then the solution
computed by the forward model must be available during the execution
of the linearised models: the adjoint and tangent linear models depend
on the forward solution.  In the tangent linear case, this is not a
major burden: the tangent linear system is block-lower-triangular,
like the nonlinear forward model, and so each tangent linear equation
can be solved immediately after the associated forward equation, with
no extra storage of the forward solutions necessary. However, as the
adjoint system is solved backwards in time, the forward solution (or
the ability to recompute it) must be available for the entire length
of the forward and adjoint solves.

In large simulations, it quickly becomes impractical to store the
entire forward solution through time at once. The alternative to
storing the forward solutions is to recompute them when necessary from
checkpoints stored during the forward run; however, a na\"ive
recomputation scheme would greatly increase the computational burden
of the problem to be solved.  Therefore, some balance of storage and
recomputation is necessary. This problem has been extensively studied
in the algorithmic differentiation
literature~\cite{griewank1992,hinze2005,wang2009,stumm2010} and
several algorithmic differentiation tools support the automatic
implementation of such a checkpointing scheme. Of particular note is
the algorithm of Griewank et al., which achieves logarithmic growth of
the storage requirements and recomputation
costs~\cite{griewank1992,griewank2000} and is provably optimal for the
case in which the number of timesteps to be performed is known in
advance~\cite{grimm1996}.

Despite these theoretical advances, implementing a checkpointing
scheme by hand is a major challenge. The complexity of programming
them inhibits their widespread use. Many adjoint models do not
implement checkpointing schemes~\cite{vidard2008}, or implement
suboptimal checkpointing schemes. In order to implement a
checkpointing scheme, the model control flow must jump between adjoint
solves and forward solves, which is difficult to achieve if the model
has a large or complex state. Instead, models may rely on temporal
interpolation schemes to approximate the forward state.  This choice
introduces approximation errors into the adjoint equations, and the
adjoint will no longer supply a gradient consistent with the discrete
forward model~\cite{tber2007}.

The main function of the annotation is to enable libadjoint to derive
the associated adjoint and tangent linear models. However, libadjoint
can also use the annotation without manipulation to assemble and solve
the original forward model. While this may not be of direct utility to
the model developer, the ability to replay the forward model means
that checkpointing schemes may be implemented entirely within
libadjoint itself. When the developer requests the
assembly of an adjoint equation from libadjoint, libadjoint will
detect any dependencies of that adjoint equation that are not
available, and will automatically recompute the relevant forward
solutions from the appropriate checkpoint, without any further
intervention by the model developer. libadjoint uses the revolve
library~\cite{griewank2000} to identify the optimal placement of
checkpoints, subject to the specified disk and memory storage
limits. If the number of timesteps is known \emph{a priori},
libadjoint uses revolve's optimal offline checkpointing
algorithm~\cite{griewank1992}; if this is not possible, libadjoint
uses revolve's online checkpointing algorithm~\cite{stumm2010}.

Once these basic checkpointing parameters have been specified, the
only change required to the model code is to add a call to the
\texttt{adj\_inc\_timestep} function at the end of the forward time
loop to indicate that the model timestep has been incremented;
libadjoint's use of revolve is fundamentally based on the timesteps of
the model, but DOLFIN currently has no native concept of time or
timestepping.  From the perspective of the model developer, the fact
that the checkpointing algorithm can be implemented entirely within
libadjoint is attractive, as it allows for the adjoints of long
forward runs to be achieved with optimal storage and recomputation
costs for almost no extra developer effort.

\subsection{Parallelism}
The algorithmic differentiation of parallel programs implemented using
OpenMP or MPI is a major research
challenge~\cite{utke2009,schanen2010,forster2011}. Even when using an
algorithmic differentiation tool to generate the majority of the
adjoint code, this challenge means that the adjoints of the
communication routines are usually written by hand. This was the
strategy used by the parallel adjoint MITgcm ocean
model~\cite{heimbach2002,heimbach2005}.

In the libadjoint context, the annotation is orthogonal to the
parallel implementation of the model: the fact that the matrices and
solutions happen to be distributed over multiple processing units is
independent of the dependency structure of the equations to be
solved. However, the callbacks supplied by the developer must be
parallel-aware: for example, the action callback of an operator must
call the relevant parallel update routine, and the transpose action must
call the adjoint of the parallel update routine. By itself, libadjoint
does not remove the need for the adjoint of the parallel communication
routines; the developer must still reverse the information flow of the
communications to implement the transpose action callbacks.

By contrast, DOLFIN handles all of these parallel communication patterns
automatically, even in the adjoint case. The high-level input to DOLFIN
contains no parallel communication calls: DOLFIN derives the correct
communication patterns at runtime.  The adjoint equations to be solved
are represented in the same UFL format as the forward model, and are
passed to the same DOLFIN runtime system. The necessary parallel
communication patterns for the adjoint equations are therefore
automatically derived in exactly the same way as the parallel
communication patterns for the forward equations, for both the MPI and
OpenMP cases~\cite[\S6.4]{logg2011}. This circumvents the need to adjoin
the parallel communication calls. Indeed, there is no parallel-specific
code in dolfin-adjoint: by operating at this high level of abstraction,
the problem of the reversal of communication patterns in the adjoint
model simply vanishes, and the adjoint model inherits the same parallel
scalability properties as the forward model. The remarkably
straightforward implementation of the parallel adjoint model is a major
advantage of adopting the combination of a high-level finite element
system and a high-level approach to deriving its adjoint.

\subsection{Matrix-free models} \label{sec:matrix-free}
In some scientific applications, the problems to be solved are so
large that it is not practical to explicitly store the matrix
associated with a single block-equation in the sense
of~\eqref{eqn:libadjoint_form}.  For example, such problems arise in
the Stokes equations for mantle convection~\cite{davies2011} and the
Boltzmann transport equations for radiation transport~\cite{ren2006}.
In such situations, matrix-free solution algorithms, algorithms that
never demand the whole matrix at once, are used instead. Instead of
assembling a matrix $M$ and passing it to the linear solver routine, a
function $f$ that computes the action $f\!\!: v \mapsto Mv$ of the
matrix on a given vector $v$ is supplied.

In general, automatically differentiating programs that use
matrix-free solvers is difficult.  If an algorithmic differentiation
tool were to be applied in a na\"ive black-box fashion, the analysis
of the tool must trace the flow of information from the independent
variables through the registration of the action function pointer into
the inner loop of the matrix-free linear solver algorithm, and
differentiate backwards through this chain. In the typical case, the
linear solver algorithm is taken from an external library, which may
be written in a different programming language, and for which the
source may not be immediately available.  Due to this difficulty, the
only previous research on algorithmically differentiating matrix-free
solvers of which the authors are aware has been limited to very
specific interfaces in PETSc that are confined to structured
meshes~\cite{hovland2002}.

However, the fact that the model takes a matrix-free approach is
ultimately an implementation detail: it is a choice made in how the
discrete problem is to be solved, but it does not change the discrete
problem or its adjoint. The difficulties faced by applying an
algorithmic differentiation tool to models that use matrix-free
solvers stem from the fact that the tool must untangle the
mathematical structure of the problem from the details of how its
solution is to be achieved; in the matrix-free case, this untangling
is particularly difficult. However, with the high-level abstraction
adopted in this work, such problems disappear: by operating at the
level of equation solves, automatically deriving the discrete adjoint
proceeds in exactly the same manner, and it does not matter whether
those equation solves happen to assemble matrices or to use
matrix-free solvers.

\subsection{Limitations}
The first major limitation of dolfin-adjoint is that all changes to object
values must happen through the DOLFIN interface. DOLFIN permits the user
to access and manipulate the raw memory addresses of function values:
however, if the user modifies the underlying array, dolfin-adjoint does
currently not identify that this has taken place, and consequently can not differentiate
through the modification. Therefore, the derived adjoint and tangent
linear models will be inconsistent with the implemented forward model.
However, the replay feature described in section \ref{sec:annotations}
mitigates this limitation somewhat: dolfin-adjoint can replay its
annotation and compare it against the values recorded in the forward
run, and automatically identify inconsistencies introduced in this way.

Another limitation of dolfin-adjoint is that it cannot
fully automate the differentiation of functionals with respect to mesh parameters,
as is typically necessary in shape optimisation. UFL is currently unable
to symbolically differentiate equation operators with respect to
the spatial coordinates; furthermore, differentiating through the mesh
generation procedure is also necessary \citep{gauger2008}. However, it is
possible to directly use the automatically computed adjoint solutions for shape
optimisation via the shape calculus approach \citep{schmidt2010,schmidt2011b},
which circumvents the need for discretely differentiating through the mesh
generation process.

\section{Examples} \label{sec:examples}

In this section, we present three numerical examples drawn from a
range of scientific applications, illustrating different
discretisations and solution strategies. These examples were run with
DOLFIN 1.0, libadjoint 0.9, and dolfin-adjoint
0.6. The software is freely available from
\url{http://dolfin-adjoint.org}.

\subsection{Cahn-Hilliard} \label{sec:cahn-hilliard}
As an initial example, the Cahn-Hilliard solver from the DOLFIN
examples collection was adjoined using the techniques described
above. The Cahn-Hilliard equation is a nonlinear parabolic
fourth-order partial differential equation used to describe the
separation of two components of a binary fluid~\cite{cahn1958,cahn1961}:
\begin{alignat}{2} \label{eqn:cahn-hilliard}
\frac{\partial c}{\partial t} - \nabla \cdot M \left( \nabla \left ( \frac{df}{dc} - \epsilon^2 \nabla^2 c \right) \right) &= 0 \quad & & \textrm{in} \ \Omega, \\
                                             M \left( \nabla \left ( \frac{df}{dc} - \epsilon^2 \nabla^2 c \right) \right) &= 0 \quad & & \textrm{on} \ \partial\Omega, \nonumber \\
                                             M \epsilon^2 \nabla c \cdot \hat{n} &= 0 \quad & & \textrm{on} \ \partial\Omega, \nonumber \\
                                                                        c(t=0)   &= c_0 \quad & & \textrm{on} \ \Omega, \nonumber
\end{alignat}
where $c$ is the unknown concentration field, $M$ and $\epsilon$ are
scalar parameters, $\hat{n}$ is the unit normal, $c_0$ is the given
initial condition, and $f = 100c^2(1 - c)^2$.

In order to solve the problem using a standard Lagrange finite element
basis, the fourth-order PDE is separated into two coupled second-order
equations:
\begin{align}
   \frac{\partial c}{\partial t} - \nabla \cdot M \nabla \mu &= 0 \quad \textrm{in} \ \Omega, \\
   \mu - \frac{df}{dc} + \epsilon^2 \nabla^{2}c &= 0 \quad \textrm{in} \ \Omega,
\end{align}
where $\mu$ is an additional prognostic variable. To solve the
problem, the equations are cast into variational form and discretised
with linear finite elements, and Crank-Nicolson timestepping
applied~\cite{crank1947}.

The example code was slightly modified for use with
dolfin-adjoint. The subclassing of objects to implement the
equation was replaced with calls to \texttt{solve}, and
\texttt{adj\_inc\_timestep} was added at the end of the timeloop as
described in section~\ref{sec:checkpointing}. These modifications were
trivial, and involved changing less than ten lines of code.

\begin{figure}
  \centering
  \subfloat[$t=0.0$]{\includegraphics[width=0.4\textwidth]{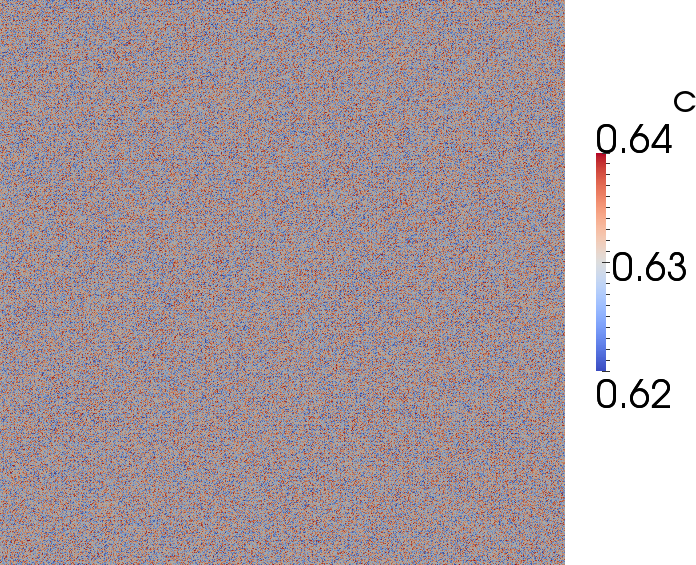}}
  \hspace{0.1\textwidth}
  \subfloat[$t=2.5\times 10^{-4}$]{\includegraphics[width=0.4\textwidth]{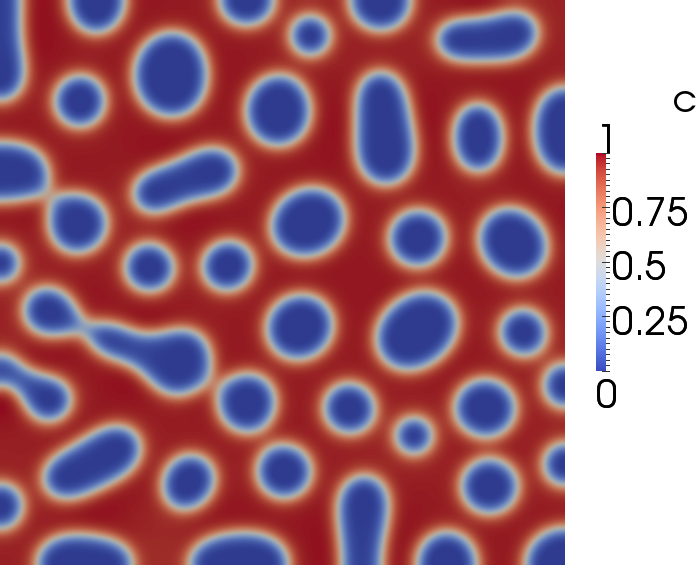}}
  \caption{The concentration $c$ of the Cahn-Hilliard problem at the initial and final time.}
  \label{fig:cahnhilliard}
\end{figure}

The functional chosen was the Willmore functional integrated
over time,
\begin{equation} \label{eqn:willmore}
\begin{split}
W(c(t), \mu(t)) &= \frac{1}{4\epsilon} \int_{t=0}^{t=T} \int_{\Omega} \left(\epsilon \nabla^2 c(t) - \frac{1}{\epsilon} \frac{df}{dc}\right)^2 \dx \dt \\
                &= \frac{1}{4\epsilon} \int_{t=0}^{t=T} \int_{\Omega} \left( -\frac{1}{\epsilon} \mu(t) \right)^2 \dx \dt,
\end{split}
\end{equation}
where $T$ denotes the final time of the simulation. This functional is
physically relevant, as it is intimately connected to the finite-time
stability of transition solutions of the Cahn-Hilliard
equation~\cite{burger2008}.  The problem was solved on a mesh of
$\Omega = [0,1]^2$ with $501,264$ vertices ($>$ 1 million degrees of
freedom), and ran for $50$ timesteps with a timestep $\Delta t=5
\times 10^{-6}$. The solution was computed on 24 cores using DOLFIN's
MPI support; both the forward and adjoint models ran in parallel with
no further modification. The solution of the concentration field at
the initial and final time are shown in figure~\ref{fig:cahnhilliard}.

To test the checkpointing implementation, libadjoint was configured to
use the multistage checkpointing algorithm of
revolve~\cite{stumm2009}, with $5$ checkpoints available in memory and
$10$ on disk. As described in section~\ref{sec:checkpointing}, the use
of this checkpointing algorithm is entirely transparent to the DOLFIN
user.

To verify the correctness of the adjoint solution, the Taylor
remainder convergence test was applied. Let $\widehat{W}(c_0)$ be the
Willmore functional considered as a pure function of the initial
condition; i.e., to evaluate $\widehat{W}(c_0)$, solve the
PDE~\eqref{eqn:cahn-hilliard} with initial condition $c_0$ to compute
$\mu(t)$, and then evaluate $W$ as in~\eqref{eqn:willmore}. The Taylor
remainder convergence test is based on the observation that, given an
arbitrary perturbation $\tilde{c}$ to the initial conditions $c_0$,
\begin{align}
\left|\widehat{W}(c_0 + h\tilde{c}) - \widehat{W}(c_0) \right| & \longrightarrow 0 \quad \textrm{at} \ O(\left|h\right|), \label{eqn:taylor_1st} \\
\intertext{but that}
\left|\widehat{W}(c_0 + h\tilde{c}) - \widehat{W}(c_0) - h\tilde{c}^T \nabla \widehat{W} \right| & \longrightarrow 0 \quad \textrm{at} \ O(\left|h\right|^2), \label{eqn:taylor_2nd}
\end{align}
where the gradient $\nabla \widehat{W}$ is computed using the adjoint
solution $z$
\begin{equation}
\nabla \widehat{W} = - \left\langle z, \frac{\partial F}{\partial c_0} \right\rangle,
\end{equation}
where $F \equiv 0$ is the discrete system corresponding
to~\eqref{eqn:cahn-hilliard}~\cite{gunzburger2003}.  This test is
extremely sensitive to even slight errors in the implementation of the
adjoint, and rigorously checks that the computed gradient is
consistent with the discrete forward model. The perturbation
$\tilde{c}$ was pseudorandomly generated, with each value uniformly
distributed in $[0, 1]$.

The results of the Taylor remainder convergence test can be seen in
table~\ref{tab:cahn-hilliard}. As expected, the convergence
orders~\eqref{eqn:taylor_1st} and~\eqref{eqn:taylor_2nd} hold,
indicating that the adjoint solution computed is correct.  Similarly,
the tangent linear model was verified, with the functional evaluated
at the final time rather than integrated. The functional gradient
in the direction of the perturbation $\tilde{c}$
is computed using the tangent linear model
\begin{equation}
\tilde{c}^T \nabla \widehat{W} = \left\langle \frac{\partial W}{\partial c_T}, \frac{\partial c_T}{\partial c_0} \tilde{c} \right\rangle,
\end{equation}
where $\frac{\partial c_T}{\partial c_0}$ is the Jacobian matrix of the final solution with respect to the initial condition.
The tangent linear model computes the term $\frac{\partial c_T}{\partial c_0} \tilde{c}$.
For the tangent linear
verification, the model was again run on 24 processors. The results of
the Taylor remainder convergence test can be seen in
table~\ref{tab:cahn-hilliard-tlm}. As expected, the theoretical
convergence orders hold, indicating that the tangent linear solution
computed is correct.
\begin{table}
\centering
\begin{tabular}{ccccc}
\toprule
$h$ & \small{$\left|\widehat{W}(\tilde{c_0}) - \widehat{W}(c_0) \right|$} & order & \small{$\left|\widehat{W}(\tilde{c_0}) - \widehat{W}(c_0) - \tilde{c_0}^T \nabla \widehat{W} \right|$} & order \\
\midrule
$1 \times 10^{-7}$ & $3.4826 \times 10^{-6}$ &  & $3.0017 \times 10^{-9}$ & \\
$5 \times 10^{-8}$ & $1.7405 \times 10^{-6}$ & $1.0006$ & $7.4976 \times 10^{-10}$ & $2.0013$ \\
$2.5 \times 10^{-8}$ & $8.7009 \times 10^{-7}$ & $1.0003$ & $1.8737 \times 10^{-10}$ & $2.0005$ \\
$1.25 \times 10^{-8}$ & $4.3499 \times 10^{-7}$ & $1.0002$ & $4.6829 \times 10^{-11}$ & $2.0004$ \\
$6.25 \times 10^{-9}$ & $2.1749 \times 10^{-7}$ & $1.0001$ & $1.1716 \times 10^{-11}$ & $1.9989$ \\
\bottomrule
\end{tabular}
\caption{The Taylor remainders for the Willmore functional
  $\widehat{W}$ evaluated at a perturbed initial condition
  $\tilde{c_0} \equiv c_0 + h\tilde{c}$, where the perturbation
  direction $\tilde{c}$ is pseudorandomly generated. All calculations were performed on the
  fine mesh with more than one million degrees of freedom. As expected, the
  Taylor remainder incorporating gradient information computed using
  the adjoint converges at second order, indicating that the
  functional gradient computed using the adjoint is correct.}
\label{tab:cahn-hilliard}
\end{table}

\begin{table}
\centering
\begin{tabular}{ccccc}
\toprule
$h$ & \small{$\left|\widehat{W}(\tilde{c_0}) - \widehat{W}(c_0) \right|$} & order & \small{$\left|\widehat{W}(\tilde{c_0}) - \widehat{W}(c_0) - \tilde{c_0}^T \nabla \widehat{W} \right|$} & order \\
\midrule
$1 \times 10^{-6}$ & 0.76441 &    & 0.03120 & \\
$5 \times 10^{-7}$ & 0.39007  & 0.9705 & 0.00773 & 2.012 \\
$2.5 \times 10^{-7}$ & 0.19698 & 0.9857 &  0.00192 & 2.006 \\
$1.25 \times 10^{-7}$ & 0.09897 & 0.9929 & $4.8005 \times 10^{-4}$ & 2.003 \\
$6.25 \times 10^{-8}$ & 0.04960 & 0.9965 &  $1.1987 \times 10^{-4}$ & 2.001 \\
\bottomrule
\end{tabular}
\caption{The Taylor remainders for the Willmore functional, with gradients computed using the tangent linear model.
  The Taylor remainders converge at second order, indicating that the tangent linear model is correct.}
\label{tab:cahn-hilliard-tlm}
\end{table}

The efficiency of the adjoint implementation was benchmarked using a
lower-resolution mesh with 40328 degrees of freedom as follows. First, the unannotated model was
run.  Then, the forward model was run again, with annotation, to
quantify the cost of annotating the forward model. Finally, the
forward and adjoint models were run together. During the forward run,
all variables were stored, and checkpointing was not used during the
adjoint run, to isolate the intrinsic cost of assembling the adjoint
system.  For each measurement, 5 runs were performed on a single
processor, and the minimum time taken for the computation was
recorded.

For this configuration, the Newton solver typically employs five
linear solves. As the adjoint replaces each Newton solve with one
linear solve, a coarse estimate of the optimal performance ratio is
1.2. The numerical results can be seen in
table~\ref{tab:cahn-hilliard-timings}. The overhead of the annotation
is less than 1\%. This overhead will further reduce with increasing mesh resolution, 
as the cost of the annotation and symbolic manipulations to derive the adjoint are independent 
of mesh size, while the costs of assembly and solves do scale with mesh size.
The adjoint model takes approximately 1.22 times the
cost of the forward model. This ratio compares very well with the
theoretical estimate: the adjoint implementation achieves almost
optimal performance.
\begin{table}
\centering
\begin{tabular}{ccc}
\toprule
       & Runtime (s) & Ratio \\
\midrule
Forward model & 103.93   &     \\
Forward model + annotation & 104.24  & 1.002 \\
Forward model + annotation + adjoint model & 127.07 & 1.22 \\
\bottomrule
\end{tabular}
\caption{Timings for the Cahn-Hilliard adjoint. The efficiency of the adjoint approaches
the theoretical ideal value of 1.2.}
\label{tab:cahn-hilliard-timings}
\end{table}

\subsection{Stokes}

\begin{figure}
  \centering
  \begin{tabular}{c}
  \includegraphics[width=12.0cm]{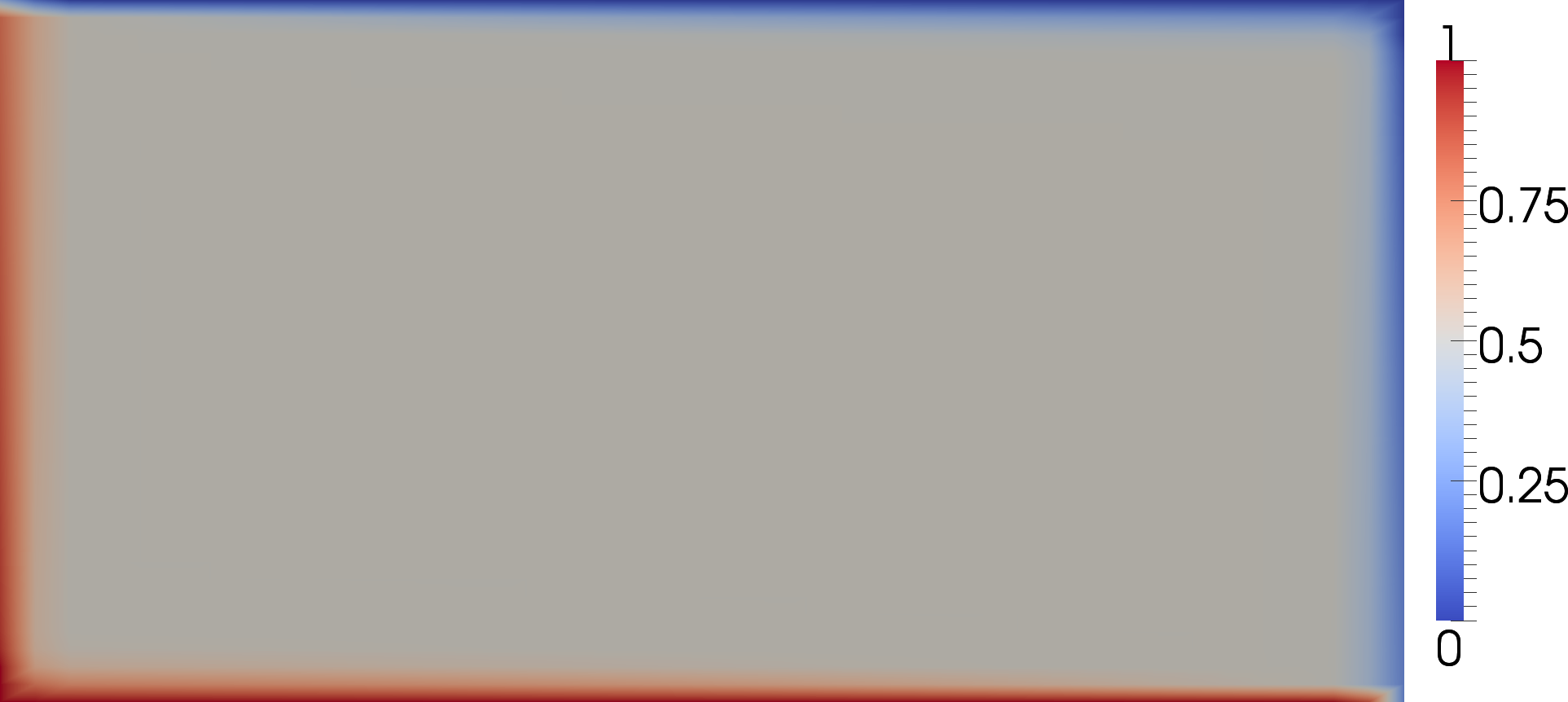}
  \\
  (a)
  \\
  \includegraphics[width=12.0cm]{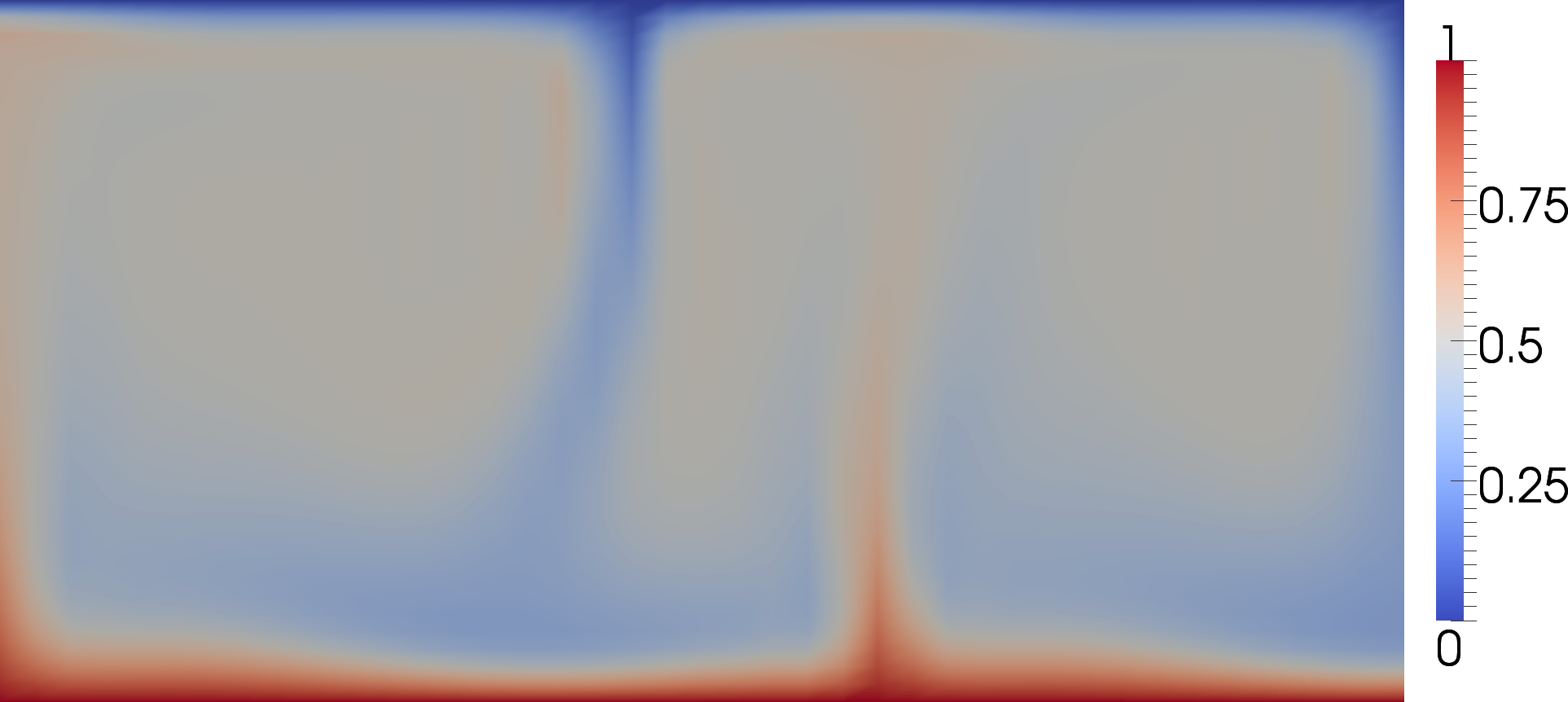}
  \\
  (b)
  \\
  \includegraphics[width=12.2cm]{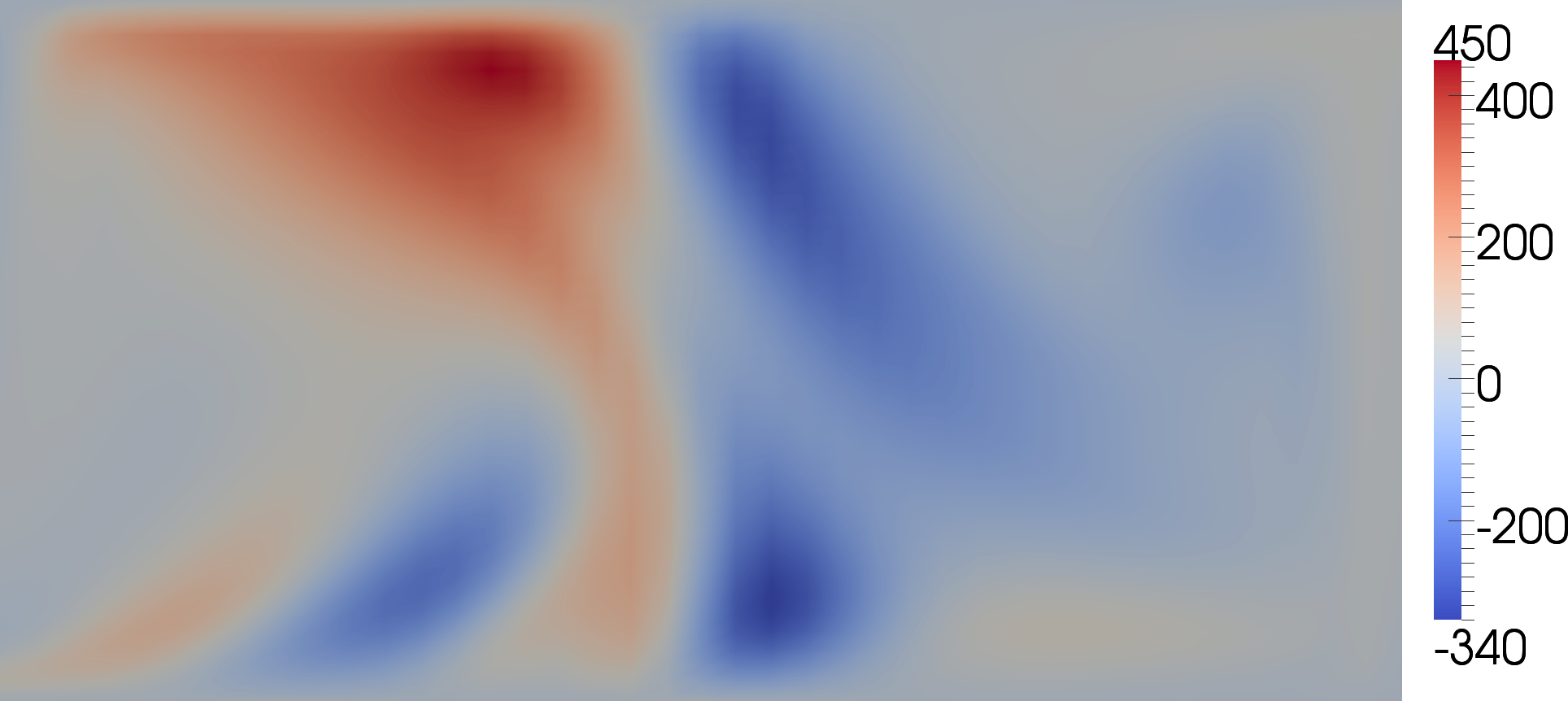}
  \\
  (c)
  \end{tabular}
  \caption{(a) The initial condition for temperature of the mantle
    convection simulation. (b) The temperature field after 200
    timesteps. Plumes are clearly visible. (c) The gradient of the
    Nusselt number with respect to the temperature initial condition,
    computed using the adjoint.}
  \label{fig:stokes}
\end{figure}

As described in section~\ref{sec:matrix-free}, the approach presented
in this paper is capable of deriving the adjoint for models that use
matrix-free solvers.  To demonstrate this capability, a matrix-free
variant of the mantle convection model presented
in~\cite{vynnytska2011} was adjoined:
\begin{align} \label{eqn:stokes}
-\nabla \cdot \sigma - \nabla p &= (\mathrm{Ra}\,T)e, \nonumber \\
              \nabla \cdot u    &= 0, \nonumber \\
\frac{\partial T}{\partial t} + u \cdot \nabla T - \nabla^2 T &= 0,
\end{align}
where $\sigma$ is the deviatoric stress tensor, $p$ is the pressure,
$\mathrm{Ra}$ is the thermal Rayleigh number, $T$ is the temperature,
$u$ is the velocity, and $e$ is a unit vector in the direction of
gravity (here the $-x_2$ direction). For the configuration reported
below, the Rayleigh number $\textrm{Ra}$ was set to $10^6$, which
yields a vigorously convective system.  For the Stokes equations, no
slip conditions are applied on the top and bottom boundaries, and a
stress-free condition is applied on the remainder of the boundary. For
the temperature equation, Dirichlet conditions are applied on the top
and bottom boundaries, while homogeneous Neumann conditions are applied
on the left and right boundaries. The initial temperature field is set
based on an analytical expression derived from boundary layer theory;
for full details, see~\cite{vynnytska2011}.

The Stokes equations were discretised using the $\mathrm{P}_2 \times
\mathrm{P}_1$ Taylor-Hood element, while the advection equation for
the temperature field was discretised using the
$\mathrm{P}_{1\mathrm{DG}}$ element. The domain $\Omega = [0, 2] \times [0, 1]$ was
discretised with 40 elements in the $x-$ and $y-$ directions, leading to a degree of freedom count of
24403. The solution of the Stokes
equations was achieved in a matrix-free manner using DOLFIN's
interfaces to the PETSc matrix-free solvers~\cite{balay2010}. Both the
forward and adjoint problems were parallelised using DOLFIN's OpenMP
support, and were run on 8 cores.

The functional taken was the Nusselt number of the temperature
evaluated at the final time
\begin{equation} \label{eqn:nusselt}
\mathrm{Nu}(T) = \int_{\Gamma_{\textrm{top}}} \frac{\partial T}{\partial x_2} \ \dx_2 \left/ \int_{\Gamma_{\textrm{bottom}}} T\ \dx_2\right.,
\end{equation}
which measures the efficiency of the convection by comparing the total
heat transferred to that transferred by thermal conduction alone. The
adjoint was computed using 30 checkpoints on disk and 30 checkpoints
in memory. The results are illustrated in figure~\ref{fig:stokes}.

The adjoint was verified using a higher-order analogue of the Taylor
remainder convergence test described in
section~\ref{sec:cahn-hilliard}.  Let $\widehat{\textrm{Nu}}$ be the
Nusselt number considered as a pure function of the initial condition
for temperature; i.e., to evaluate $\widehat{\textrm{Nu}}(T_0)$, solve
the PDE~\eqref{eqn:stokes} with initial condition $T_0$ to compute the
final temperature $T$, and then evaluate $\textrm{Nu}(T)$ as
in~\eqref{eqn:nusselt}. The test is based on the observation that,
given an arbitrary perturbation $\tilde{T}$ to the initial conditions
$T_0$,
\begin{align} \left|\widehat{\textrm{Nu}}(T_0 + \frac{h}{2}\tilde{T}) - \widehat{\textrm{Nu}}(T_0 - \frac{h}{2}\tilde{T}) \right| & \longrightarrow 0 \quad \textrm{at} \ O(\left|h\right|), \label{eqn:taylor_high_1st} \\
\intertext{but that}
\left|\widehat{\textrm{Nu}}(T_0 + \frac{h}{2}\tilde{T}) - \widehat{\textrm{Nu}}(T_0 - \frac{h}{2}\tilde{T}) - h\tilde{T}^T \nabla \widehat{\textrm{Nu}} \right| & \longrightarrow 0 \quad \textrm{at} \ O(\left|h\right|^3), \label{eqn:taylor_high_3rd}
\end{align}
where the gradient $\nabla \widehat{\textrm{Nu}}$ is computed using
the adjoint solution $z$
\begin{equation}
\nabla \widehat{\textrm{Nu}} = - \left\langle z, \frac{\partial F}{\partial T_0} \right\rangle,
\end{equation}
where $F \equiv 0$ is the discrete system corresponding to
\eqref{eqn:stokes}. The higher-order version of the Taylor remainder
convergence test was used because it reaches the asymptotic region
more rapidly, which is useful for strongly nonlinear problems such as
this. The results of the Taylor remainder convergence test can be
seen in table~\ref{tab:stokes}. As expected, the convergence
orders~\eqref{eqn:taylor_high_1st} and~\eqref{eqn:taylor_high_3rd}
hold, indicating that the adjoint solution computed is correct.
\begin{table}
\centering
\begin{tabular}{ccccc}
\toprule
$h$ & \small{$\left|R\,\right|$} & order & \small{$|R - h\tilde{T}^T \nabla \widehat{\textrm{Nu}} |$} & order \\
\midrule
$7.5 \times 10^{-3}$ & 0.14568 &    & $3.6260 \times 10^{-2}$ & \\
$3.75 \times 10^{-3}$ & 0.05905  & 1.302 & $4.3465 \times 10^{-3}$ & 3.06 \\
$1.875 \times 10^{-3}$ & 0.02787 & 1.083 &  $5.1659 \times 10^{-4}$ & 3.07 \\
$9.375 \times 10^{-4}$ & 0.01373 & 1.021 & $5.2311 \times 10^{-5}$ & 3.30 \\
\bottomrule
\end{tabular}
\caption{The Taylor remainder $R \equiv \widehat{\textrm{Nu}}(T_0 +
  \frac{h}{2}\tilde{T}) - \widehat{\textrm{Nu}}(T_0 -
  \frac{h}{2}\tilde{T})$ for the Nusselt functional
  $\widehat{\textrm{Nu}}$ evaluated at a perturbed initial condition,
  where the perturbation direction $\tilde{T}$ is the vector of all
  ones. All calculations were performed on the mesh with 24403 degrees
  of freedom. The third-order convergence of the Taylor remainders indicates that the adjoint
  is correct.}
\label{tab:stokes}
\end{table}

The efficiency of the adjoint implementation was benchmarked using the
same procedure as described in section~\ref{sec:cahn-hilliard}, again
using a lower-resolution configuration of 1603 degrees of freedom. The results can be seen in
table~\ref{tab:stokes-timings}. The performance overhead of solving the linear systems
matrix-free is significant, and these timings should not be taken as representative
of the potential performance of the modelling system. The overhead of the annotation is
extremely small, within the distribution of timings of the unannotated
run. The adjoint model takes approximately 86\% of the cost of the
forward model. In this case, the forward model performs two Picard
iterations per timestep, each of which induce a corresponding linear
solve in the adjoint equations.  Therefore, a na\"ive estimate of the
ideal theoretical efficiency is 2.  On investigation, the proximate
cause of the adjoint run being cheaper than the forward run was that
the matrix-free linear solvers happened to converge more quickly
during the adjoint run.
\begin{table}
\centering
\begin{tabular}{ccc}
\toprule
       & Runtime (s) & Ratio \\
\midrule
Forward model & 96.03   &     \\
Forward model + annotation & 96.05  & 1.0 \\
Forward model + annotation + adjoint model & 178.75 & 1.86 \\
\bottomrule
\end{tabular}
\caption{Timings for the Stokes mantle convection adjoint. The efficiency of the adjoint
exceeds the theoretical ideal value of 2, as the adjoint linear solves happen to converge
faster.}
\label{tab:stokes-timings}
\end{table}

\subsection{Viscoelasticity}

Most biological tissue responds in a viscoelastic, rather than purely
elastic, manner. As a final example, we
consider a nontrivial discretisation of a viscoelastic model for the
deformation and stress development in the upper part of the spinal
cord under pressure induced by the pulsating flow of cerebrospinal
fluid \cite{greitz2006}.

The {Standard Linear Solid} viscoelastic model equations can be
phrased~\cite{rognes2010b} as: find the Maxwell stress tensor
$\sigma_0$, the elastic stress tensor $\sigma_1$, the velocity $v$ and
the vorticity $\gamma$ such that
\begin{equation}
  \label{eq:visco}
  \begin{split}
    A^0_1 \frac{\partial}{\partial t} \sigma_0 + A^0_0 \sigma_0
    - \nabla v + \gamma &= 0,\\
    A^1_1 \frac{\partial}{\partial t} \sigma_1
    - \nabla v  + \gamma &= 0, \\
    \nabla \cdot (\sigma_0 + \sigma_1) &= 0 , \\
    \mathrm{skw}(\sigma_0 + \sigma_1) &= 0,
  \end{split}
\end{equation}
for $(t; x, y, z) \in (0, T] \times \Omega$. Here, $A^0_1$, $A^0_0$,
$A^1_1$ are fourth-order compliance tensors, the divergence and
gradient are taken row-wise, and $\mathrm{skw}$ denotes the
skew-symmetric component of a tensor field. The total stress tensor
$\sigma$ is the sum of the Maxwell and elastic contributions.  In
the isotropic case, each of the compliance tensors $A^0_0, A^0_1,
A^1_1$ reduce to two-parameter maps:
\begin{equation}
  A^i_j = \left( C_j^{i} \right)^{-1}, \quad
  C_j^{i} \varepsilon = \mu^i_j \varepsilon
  + \lambda^i_j \mathrm{tr}(\varepsilon) I,
\end{equation}
where $\mu^i_j, \lambda^i_j$ are positive Lam\'e parameters. The
system is closed by initial conditions for the Maxwell stress,
essential boundary conditions for the velocity, and traction boundary
conditions for the total stress $\sigma \cdot \hat{n}$, where
$\hat{n}$ is the outward normal on the domain boundary.  The cord was
kept fixed at the top and bottom, and the parameters were set to
$\mu^0_0 = 37.466$, $\lambda^0_0 = 10^4$, $\mu^0_1 = 4.158$,
$\lambda^0_1 = 10^3$, $\mu^1_1 = 2.39$, $\lambda^1_1 = 10^3$ (kPa).
The traction boundary condition was set to
\begin{equation*}
\sigma \cdot \hat{n} = - p \hat{n},
\end{equation*}
where $p$ is a periodically varying pressure modelled as $p(t; x, y,
z) = a \sin(2 \pi t)(171 - 78)^{-1} (z - 78)$ (kPa), where $a = 0.05$
is the amplitude of the pressure.

The discretisation of~\eqref{eq:visco} is performed using the
(locking-free) scheme introduced in~\cite{rognes2010b}, allowing for
direct approximation of the stresses, while enforcing the symmetry of
the total stress weakly.  The temporal discretisation is carried out
via a two-step $\mathrm{TR-BDF_2}$ scheme; that is, a Crank-Nicolson
step followed by a $2$-step backward difference scheme, while the
spatial, mixed finite element discretisation is based on seeking
approximations $\sigma_0(t), \sigma_1(t), v(t), \gamma(t)$ in the
space
\begin{equation}
  Z = \mathrm{BDM}_1^3 \times \mathrm{BDM}_1^3 \times \mathrm{P}_{0\mathrm{DG}}^3
  \times \mathrm{P}_{0\mathrm{DG}}^3.
\end{equation}
where $\mathrm{BDM}_1$ denotes the lowest-order
$H(\mathrm{div})$-conforming Brezzi-Douglas-Marini elements and
$\mathrm{P}_{0\mathrm{DG}}$ denotes piecewise
constants. See~\cite{rognes2010b} for more details.

Abnormal stress conditions in the interior of the spinal cord may be
of biomedical interest~\cite{levine2004,heiss1999}.  In particular, we
focus on the contribution of the Maxwell stress tensor in the
horizontal plane at the final time $T$:
\begin{equation}
  \label{eq:visco:functional}
  J = \int_{\Omega} \left ( \sigma_0(T) \cdot e_z \right ) ^2 \mathrm{d} x,
\end{equation}
where $e_z$ denotes a unit vector in the $z$-direction.

\begin{figure}
  \centering
  {\includegraphics[width=0.3\textwidth]{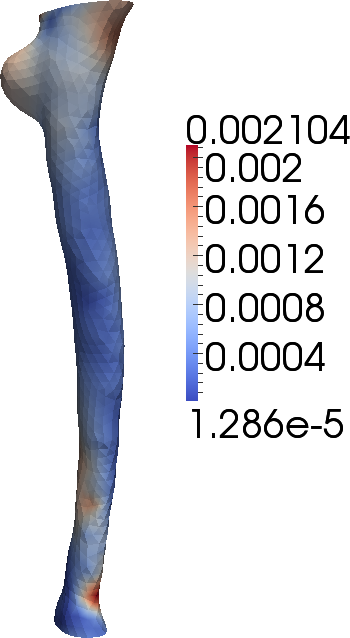}}
  \caption{The magnitude of the Maxwell stress in the horizontal
    plane at $t = 1.25$.}
  \label{fig:viscoelastic}
\end{figure}

The problem was solved on a tetrahedral mesh generated from
patient-specific imaging data, yielding a total of $879204$ degrees
of freedom, for $t \in [0, 1.25]$ with a time step of $\Delta t =
0.01$. As the system is linear, the matrices corresponding to each of
the steps in the $\mathrm{TR-BDF_2}$ scheme were pre-assembled outside
of the timeloop and their LU factorisations were cached. This
optimisation is recognised by dolfin-adjoint, which then
applies the analogous factorisation strategy to the corresponding
adjoint solve. The Maxwell stress in the horizontal plane at $t =
1.25$ is illustrated in figure~\ref{fig:viscoelastic}.
\begin{figure}
  \centering
  \subfloat[]{\includegraphics[width=0.49\textwidth]{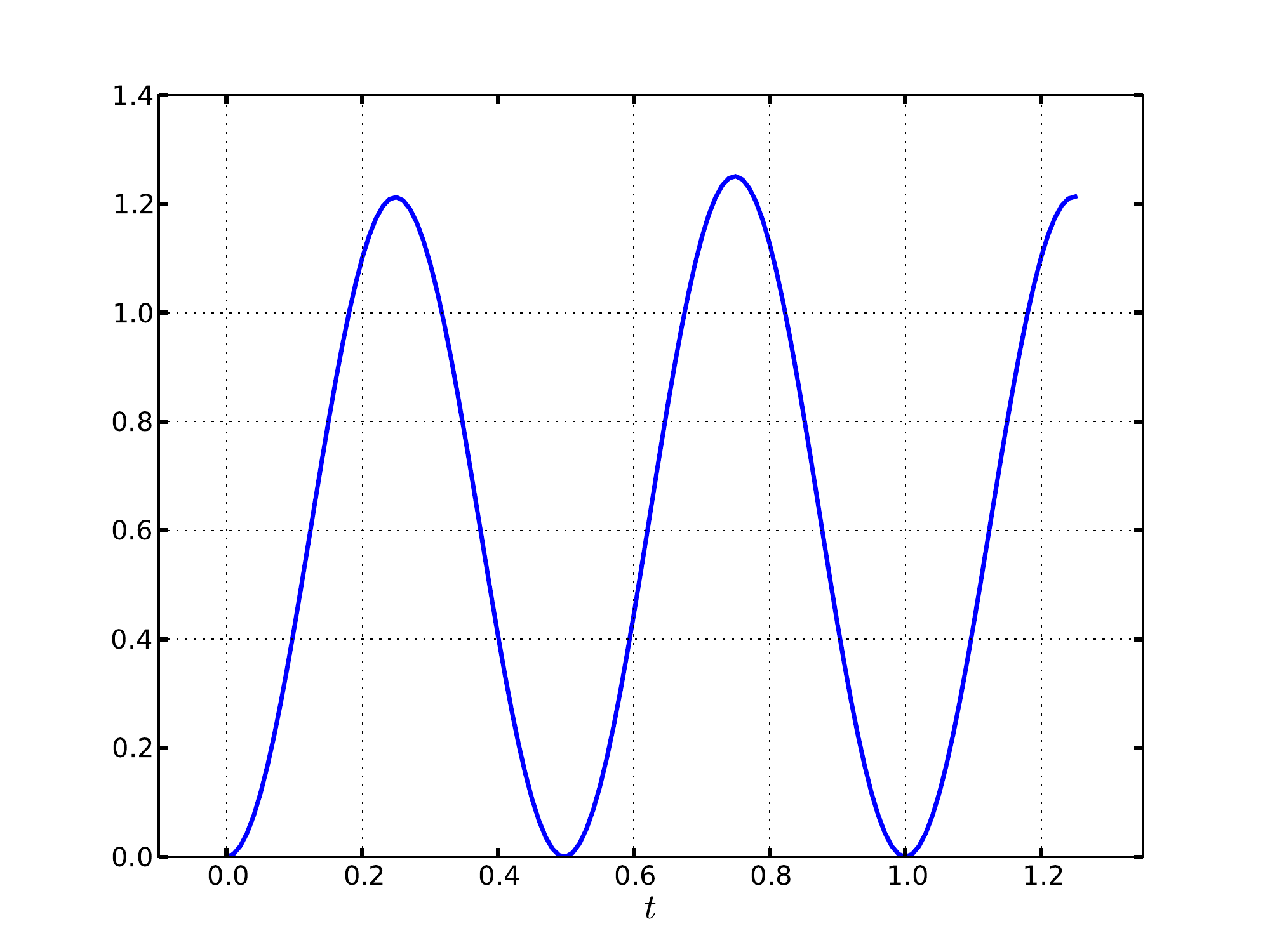}}
  \subfloat[]{\includegraphics[width=0.49\textwidth]{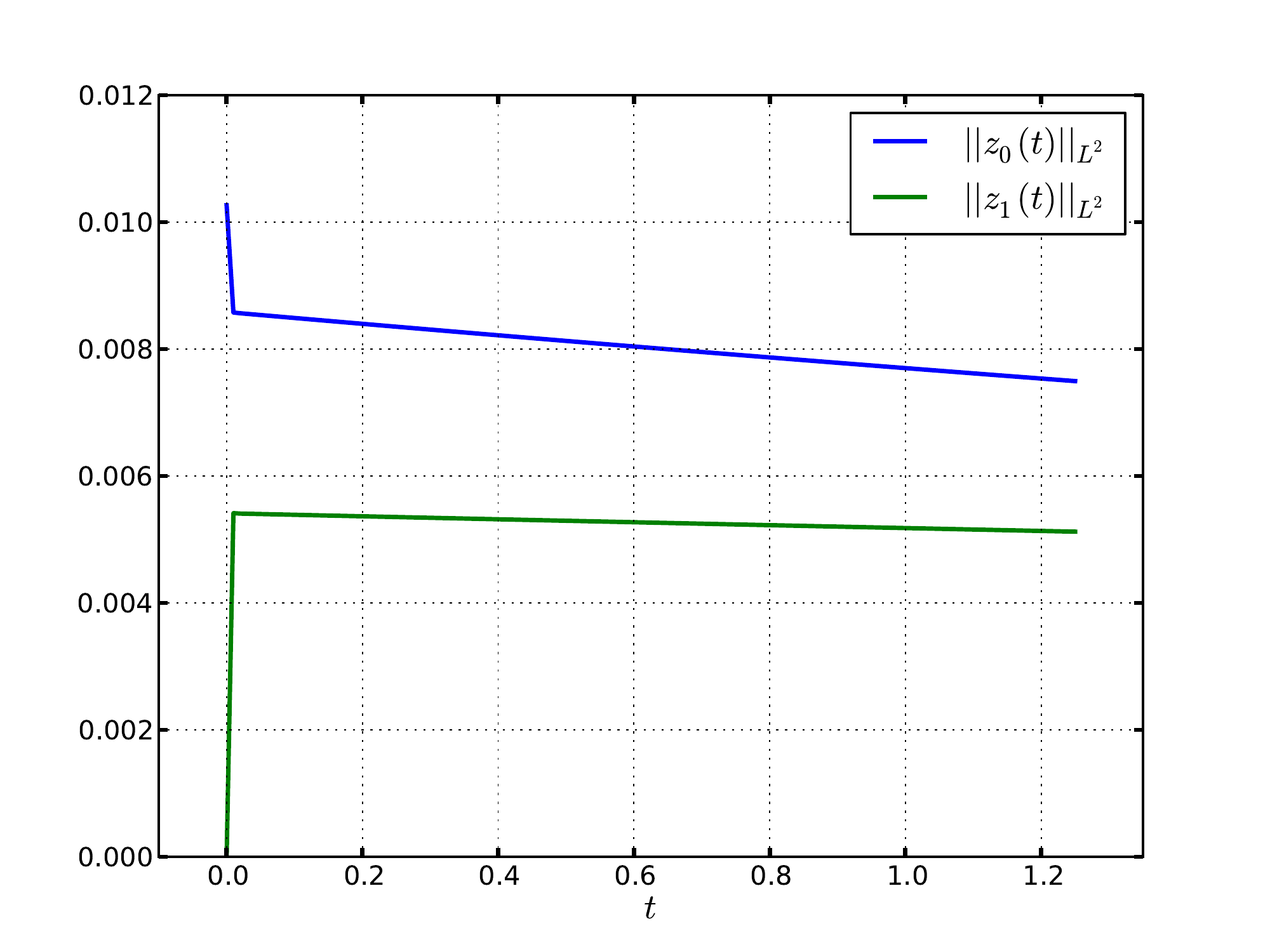}}
  \caption{(a) $L^2$-norm squared of the Maxwell stress in the
    horizontal plane versus time; (b) The $L^2$-norm of the adjoint
    Maxwell stress tensor $z_0$ and the adjoint elastic stress tensor
    $z_1$ versus time.}
  \label{fig:viscoelastic:norms}
\end{figure}

To verify the correctness of the adjoint solution, the Taylor
remainder convergence test was applied. Let $\widehat{J}(a)$ be the
Maxwell stress functional considered as a pure function of the
amplitude of the applied pressure; i.e., to evaluate $\widehat{J}(a)$,
solve the PDE~\eqref{eq:visco} with pressure amplitude $a$ to compute
$\sigma_0(T)$, and then evaluate $J$ as
in~\eqref{eq:visco:functional}.  The results of the Taylor remainder
convergence test can be seen in table~\ref{tab:visco}. As expected,
the convergence orders~\eqref{eqn:taylor_1st}
and~\eqref{eqn:taylor_2nd} hold, indicating that the adjoint solution
computed is correct.
\begin{table}
\centering
\begin{tabular}{ccccc}
\toprule
$\delta a$ & \small{$\left|\widehat{J}(a + \delta a) - \widehat{J}(a) \right|$} & order & \small{$\left|\widehat{J}(a + \delta a) - \widehat{J}(a) - {\nabla \widehat{J}}\cdot \delta a \right|$} & order \\
\midrule
$0.05$ & $9.1012 \times 10^{-3}$ &  & $3.0337 \times 10^{-3}$ & \\
$0.025$ & $3.7921 \times 10^{-3}$ & 1.2630 & $7.58417 \times 10^{-4}$ & 2.0000 \\
$0.0125$ & $1.7064 \times 10^{-3}$ & 1.1520 & $1.8959 \times 10^{-4}$ & 2.0000 \\
$6.25 \times 10^{-3}$ & $8.0583 \times 10^{-4}$ & 1.0824 & $4.7397 \times 10^{-5}$ & 2.0001 \\
$3.125 \times 10^{-3}$ & $3.9106 \times 10^{-4}$ & 1.0430 & $1.1848 \times 10^{-5}$ & 2.0001 \\
\bottomrule
\end{tabular}
\caption{The Taylor remainders for the functional given
  by~\eqref{eq:visco:functional}. All calculations were performed on the mesh
  with $879204$ degrees of freedom. The convergence of the Taylor remainders indicates
  that the adjoint is correct.}
\label{tab:visco}
\end{table}

The efficiency of the adjoint implementation was benchmarked using the
same procedure as described in section~\ref{sec:cahn-hilliard}, again
using a lower-resolution configuration of 86976 degrees of freedom. The results can be seen in
table~\ref{tab:viscoelasticity-timings}. As the problem is linear, the
theoretical estimate of the ideal efficiency ratio is 2, which is
approximately achieved by the implementation presented here.
\begin{table}
\centering
\begin{tabular}{ccc}
\toprule
       & Runtime (s) & Ratio \\
\midrule
Forward model & 119.93   &     \\
Forward model + annotation & 120.24  & 1.002 \\
Forward model + annotation + adjoint model & 243.99 & 2.029 \\
\bottomrule
\end{tabular}
\caption{Timings for the viscoelasticity adjoint. The efficiency of the adjoint
approaches the theoretical ideal value of 2.}
\label{tab:viscoelasticity-timings}
\end{table}

\section{Conclusion}
\label{sec:conclusion}

Naumann's recent book on algorithmic differentiation
states~\cite[pg.~\emph{xii}]{naumann2011}:
\begin{quote}
  [T]he automatic generation of optimal (in terms of robustness and
  efficiency) adjoint versions of large-scale simulation code is one
  of the great open challenges in the field of High-Performance
  Scientific Computing.
\end{quote}
The framework presented here, dolfin-adjoint, provides a
robust and efficient mechanism for automatically deriving adjoint and
tangent linear models of a wide variety of finite element models
implemented in the Python interface to the DOLFIN library. Only
minimal changes are required to adapt such a forward model for use
with dolfin-adjoint. The adjoint model draws on the
advantages of libadjoint to deliver optimal checkpointing strategies,
and inherits the seamless parallelism of the FEniCS framework. The
numerical results obtained demonstrate optimal efficiency in the
adjoint model.

The approach employed in dolfin-adjoint is analogous but fundamentally
different to that adopted by algorithmic differentiation: by operating
at a higher level of abstraction, a much greater degree of automation
and efficiency has been achieved. The framework presented here enables
adjoint models to be derived automatically, reliably and robustly,
relieving the model developers of the adjoint development task.

\bibliographystyle{siam}
\bibliography{literature}

\end{document}